\documentclass[showpacs,preprint,preprintnumbers,amsmath,amssymb]{revtex4}
\begin{document}

\preprint{APS/123-QED}
\title{Nuclear matter incompressibility coefficient in relativistic and
nonrelativistic microscopic models}

\author{B. K. Agrawal,  S. Shlomo,  and V. Kim Au}
\address{Cyclotron Institute, Texas A\&M University,
College Station, TX 77843, USA}

\begin{abstract}
We systematically analyze the recent claim that nonrelativistic and
relativistic mean field (RMF) based random phase approximation (RPA)
calculations for the centroid energy $E_0$ of the isoscalar giant
monopole resonance yield for the nuclear matter incompressibility
coefficient, $K_{nm}$, values which differ by about $20\%$. For an
appropriate comparison with the RMF based RPA  calculations, we obtain
the parameters of the  Skyrme force used in the nonrelativistic model
by adopting the same procedure as employed in the determination of the
NL3 parameter set of the effective Lagrangian used in the RMF model. Our
investigation suggests that the discrepancy between the values of $K_{nm}$
predicted by the relativistic and nonrelativistic models is significantly
less than $20\%$.
\end{abstract} 
 \maketitle

\section{Introduction}
The nuclear matter incompressibility coefficient  $K_{nm}$  plays an
important role in understanding  a wide variety of phenomena ranging from
giant resonances in finite nuclei to heavy-ion collisions and supernova
explosions. The measurement  of the centroid energy of the isoscalar
giant monopole resonance (ISGMR)  provides a very sensitive method
to determine the value of $K_{nm}$. Over the last  couple of decades,
attempts have been made to measure very accurately the value of the ISGMR
centroid energy, $E_0$. The recent experimental data \cite{Youngblood99}
for the $E_0$ in heavy nuclei are accurate enough  to provide  unambiguous
information about $K_{nm}$.  However, the theoretical scenario   of the
$K_{nm}$ still remains unclear.

In the  past, the value of $K_{nm}$ was determined using a macroscopic
approach which relies on the liquid drop type of expansion for the
breathing mode restoring force. It was shown that equally good fits
can be obtained with values of $K_{nm}$ ranging from $100 - 400$ MeV
\cite{Pearson91,Shlomo93}.  In other words, the liquid drop approach can
not constrain the value of $K_{nm}$ better than $50\%$.  On the other
hand, the microscopic determination of $K_{nm}$ using the Hartree-Fock
based random phase approximation (HF-RPA) has undergone a significant
improvement over time.  The earlier nonrelativistic HF$-$RPA calculations
carried out using the  Skyrme interaction, which nicely reproduced the
gross properties of nuclei (such as nuclear binding energy and charge
radii) and the available data on isovector giant dipole resonance and
isoscalar giant quadrupole resonance, yielded values of about 370 MeV
for the $K_{nm}$. With these interactions, the ISGMR in $^{208}{\rm
Pb}$ was predicted to be located at an excitation energy of about
18 MeV.  The discovery of ISGMR in $^{208}{\rm Pb}$ at an excitation
energy of 13.7 MeV \cite{Marty75} led to the modification of the Skyrme
interaction. Until today,    the nonrelativistic calculations with Skyrme
\cite{Colo92,Hamamoto97} and Gogny \cite{Blaizot95} interactions predict
a  value of $K_{nm}$ in the range of $210 - 220$ MeV. We remark here  that
the long standing problem of the conflicting results deduced for $K_{nm}$
from data on the isoscalar giant dipole resonance (ISGDR) and data on
the ISGMR was explained recently by Shlomo and Sanzhur \cite{Shlomo02}
as being due to a missing strength in the experimental data for the high
energy region of the ISGDR.

The relativistic mean field based RPA  (also referred to as RRPA)
calculations, with the contribution from negative-energy sea not
included, yielded for $K_{nm} $ a value in the range of  $280 - 350$
MeV \cite{Ma97}.  Recent RRPA calculations \cite{Ma01,Vretenar03}, with
the inclusion of negative-energy states in the response function,  yield a
value of $K_{nm}$  between $250-270$ MeV.  The discrepancy of about $20\%$
in the value of  $K_{nm}$  obtained from relativistic and nonrelativistic
models  is quite significant in view of the accuracy of the  experimental
data currently   available on the ISGMR centroid energies. In recent
studies \cite{Giai01,Niksic02} it has been claimed that these pronounced
differences are due to the model dependence of $K_{nm}$. On the other
hand, it has been pointed out  in Ref. \cite{Piekarewicz02} that the
differences in the values of $K_{nm}$ obtained in  the relativistic
and the nonrelativistic models can be attributed, at least in part,
to the differences in the density dependence of the symmetry energy in
these models.  However, in Ref. \cite{Piekarewicz02}, the analysis was
restricted to a single nucleus $^{208}Pb$ and the interaction parameters
for the several families of the effective Lagrangian considered
were fitted only to the empirical values of saturation density,
binding energy per nucleon in symmetric nuclear matter and the charge
radius of the $^{208}Pb$ nucleus.  It may be pointed out here that in
Ref. \cite{Serr81}  a reasonable value of $E_0$ for the $^{208}{\rm Pb}$
nucleus was obtained using an effective force with $K_{nm} =400$ MeV.
But, the same effective force overestimated the value of $E_0$ in case
of the $^{90}{\rm Zr}$ nucleus.  Moreover, it has been suggested in
Ref. \cite{Serr81} that a wide range of combinations of bulk, surface
and asymmetry contributions to the finite nucleus compressibility can fit
the energy of the ISGMR in medium to heavy nuclei. This implies that for
a meaningful informations about the discrepancy between the relativistic
RRPA and  the nonrelativistic HF-RPA calculations, one must compare the
results obtained from these models for several nuclei.

It is noteworthy that in a crude approximation, the uncertainty of
about $20\%$ in the values of $K_{nm}$ is tantamount to an uncertainty
of $10\%$ in the value of  $E_0$. This is because, in a semi-classical
approach, $E_0\propto \sqrt{K_{nm}}$.  We have shown very recently
\cite{Agrawal03} that the calculated  value of $E_0$ can deviate
by about $5\%$ if the particle-hole space is quite limited and/or
self-consistency is not properly maintained.  We note that in the
published literature, the calculated values of $E_0$ for $^{208}{\rm
Pb}$, obtained for the same interaction, differs by up to 0.3 MeV
\cite{Giai01,Colo99,Colo01}. Therefore, appropriate comparison for the
values of $K_{nm}$ obtained from  the different models is possible only
when all the calculations are performed with the same procedure and
numerical accuracy.

In this work we take a close look at the issue of the model dependence
of the nuclear matter incompressibility coefficient derived from the
ISGMR centroid energy.  Toward this purpose, we generate different
parameter sets for the Skyrme interaction and perform highly accurate
calculations for the ISGMR strength function for several nuclei using
the HF based RPA approach.  For the sake of true comparison,  the
calculations using different  parameter sets of Skyrme interaction are
performed following exactly the same numerical procedure.  The values of
the Skyrme parameters are obtained by a least square fit to exactly the
same experimental data for the nuclear binding energies, charge radii
and neutron radii as adopted in Ref. \cite{Lalazissis97} in determining
the NL3 parameter set for an effective Lagrangian used in the RMF model.
We find that the model dependence is rather weak and the differences in
the values of $K_{nm}$ in the relativistic and the nonrelativistic models
essentially arise from the differences in the nuclear matter properties,
in particular, in the values of the symmetry energy coefficient ($J$),
associated with these models.

\section{Formalism}
In a self-consistent HF-RPA calculation \cite{Bertsch75}, one starts by
adopting a specific effective nucleon-nucleon interaction, $V_{12}$. In
this work we shall use the Skyrme type interaction of the
form \cite{Vautherin72},
\begin{eqnarray}
V_{12}&=& t_0\left (1+x_0 P_{12}^\sigma\right )\delta({\bf r}_1-{\bf r}_2)
+\frac{1}{2}t_1\left (1+x_1 P_{12}^\sigma\right ) \nonumber\\
&&\times \left[\overleftarrow{k}_{12}^2\delta({\bf r}_1-{\bf r}_2)+\delta({\bf r}_1-{\bf r}_2)\overrightarrow{k}_{12}^2\right] 
+t_2\left (1+x_2 P_{12}^\sigma\right )\overleftarrow{k}_{12}\delta({\bf r}_1-{\bf r}_2)\overrightarrow{k}_{12} \nonumber\\
&&+ \frac{1}{6}t_3\left (1+x_3 P_{12}^\sigma\right )\rho^\alpha\left(\frac{{\bf r}_1+{\bf r}_2}{2}\right )
\delta({\bf r}_1-{\bf r}_2) \nonumber\\
&&+iW_0\overleftarrow{k}_{12} \delta({\bf r}_1-{\bf r}_2)(\overrightarrow{\sigma_1}+
\overrightarrow{\sigma_2})\times \overrightarrow{k}_{12}
\end{eqnarray}
where, $P_{12}^\sigma$ is the spin exchange operator,$\overrightarrow{\sigma}_i$ is the Pauli spin operator,
$\overrightarrow{k}_{12} = -i( \overrightarrow{\nabla}_1-\overrightarrow{\nabla}_2)/2$
 and 
$\overleftarrow{k}_{12} = -i(\overleftarrow{\nabla}_1-\overleftarrow{\nabla}_2)/2\, .$
Here,  right and left arrows indicate that the momentum operators act on
the right and  on the left ,  respectively.  The parameters of the Skyrme
force are obtained by fitting the HF results to the experimental data
for the bulk properties of finite nuclei.  Once the HF equations are
solved using an appropriate  parameter set for the Skyrme interaction,
then one obtains the RPA Green's function \cite{Bertsch75}
\begin{equation}
G = G_0 (1+ V_{p-h}G_0)^{-1}\, ,
\end{equation}
where, $V_{p-h}$ is the particle-hole ($p-h$) interaction consistent with $V_{12}$ and 
$G_0$ is the free $p-h$ Green's function . For the single-particle operator  
\begin{equation}
F=\sum_{i=1}^{A} f({\bf r}_i),
\label{operator-F}
\end{equation}
the strength function  is given by 
\begin{equation}
S(E)=\sum_n\left | \left <0\left |F\right |n \right >\right |^2 \delta \left(E-E_n\right )=
\frac{1}{\pi}{ Im}\left [ Tr \left (fGf\right )\right ].
\label{strength}
\end{equation}

The steps involved in the relativistic mean field based RPA calculations
are analogous  to  those described above for the nonrelativistic
HF-RPA approach. However, the nucleon-nucleon interaction in case of
relativistic mean field models are generated through the exchange of
various mesons. An effective Lagrangian which represent a system of
interacting nucleons looks like \cite{Lalazissis97},
\begin{eqnarray}
\label{lagrangian}
{\cal L}&=&\bar{\psi}\left (\gamma\left (i\partial - g_\omega\omega -
g_\rho \overrightarrow{\rho}\overrightarrow{\tau} -e A\right )-m -
g_\sigma \sigma\right ) \psi +\frac{1}{2}\left (\partial \sigma\right )^2\nonumber\\
&& - U(\sigma) - \frac{1}{4}\Omega_{\mu\nu} \Omega^{\mu\nu}+ \frac{1}{2} m_\omega^2 \omega^2 -
\frac{1}{4}\overrightarrow{R}_{\mu\nu}\overrightarrow{R}^{\mu\nu}\nonumber\\
&&+ \frac{1}{2} m_\rho^2\overrightarrow{\rho}^2 - \frac{1}{4}F_{\mu\nu}F^{\mu\nu}
\end{eqnarray}
which contains nucleons $\psi$ with mass $m$; $\sigma$, $\omega$, $\rho$
mesons; the electromagnetic fields; and non-linear self-interactions of
the $\sigma$ field, 
\begin{equation}
\label{u-sig}
U(\sigma) = \frac{1}{2}m_\sigma^2\sigma^2 + \frac{1}{3}
g_2\sigma^3+\frac{1}{4}g_3\sigma^4.
\end{equation}
The Lagrangian parameters are  usually obtained, as in the case
of nonrelativistic mean field calculations, by fitting procedure to
some bulk properties of a set of spherical nuclei \cite{Reinhard89}.
The values of various coupling constants and the meson masses  appearing
in Eqs.  (\ref{lagrangian}) and (\ref{u-sig})  for the most widely used
parameter set NL3  are $ m_\sigma = 508.194$ MeV, $m_\omega = 782.501$
MeV, $m_\rho= 763.000$ MeV, $g_\sigma = 10.217$, $g_\omega = 12.868$,
$g_\rho=4.474$, $g_2 = -10.431$ fm$^{-1}$ and $g_3 = -28.885$.

\section{Results and discussions}
\subsection{HF  results}
In the present work, for an appropriate comparison, we carry out
a least square fit  to the same experimental data for the nuclear
binding energies, charge radii and neutron radii  used in Ref.
\cite{Lalazissis97} to  obtain the NL3 parameter set.  Furthermore,
we  deal with  the centre of mass correction to the total binding
energy, finite size effects of the proton and the Coulomb energy  in
the way similar to that employed in determining the NL3 parameter set
in Ref. \cite{Lalazissis97}.  It may be pointed out that pairing is not
included in our HF calculations and  instead of the 10 nuclei considered
in Ref. \cite{Lalazissis97}, we consider 7 closed shell nuclei. The
open shell nuclei  $^{58}{\rm Ni}$, $^{124}{\rm Sn}$ and $^{214}{\rm Pb}$ are
excluded from our least square fit.  We also ignore the proton and the
neutron pairing gaps in case of $^{90}{\rm Zr}$ and $^{116}{\rm Sn}$
nuclei, respectively. However, we have verified that if we increase
the  error bars for the experimental data on these two nuclei in order
to compensate  for the pairing, we find that the values of the Skyrme
interaction  parameters remains practically unaltered.

Since, the main objective of this paper is to delineate the differences in
the value of $K_{nm}$ predicted by the relativistic and nonrelativistic
mean field based RPA calculations, we generate a Skyrme interaction
having $K_{nm}$ and $J$ very close to  those associated with the NL3
parameter set, i.e., 271.76 and 37.4 MeV,  respectively. Also, as most
of the calculation claiming the model dependence are restricted to the
ISGMR centroid energy for the  single heavy nucleus $^{208}{\rm Pb}$, we
generate a parameter set  by demanding  a very high accuracy for the root
mean square charge radius of $^{208}{\rm Pb}$. We denote this parameter
set as SK272. In addition, we also obtain a parameter set SK255 having
characteristics very much similar to  the SK272 parameter set,  but, $K_{nm}$
is taken to be 255 MeV.

In Table \ref{parameters} we give the values of the Skyrme parameters
SGII together with the new parameter sets SK272 and SK255.  In Table
\ref{n-matter} we compare the nuclear matter properties for the SK272 and
SK255  interactions with the corresponding ones obtained from the  NL3 and
SGII interactions. The quantities $\rho_0$, $m^*/m$ and $L$ in this table
denote  the saturation density, effective nucleon mass and the slope of
the symmetry energy coefficient ($L=3\rho_0\,dJ/d\rho_0$), respectively.
In column 3 of Table \ref{f-nuclei} we have given the experimental
data for the  total binding energy, $E$,  charge radii, $r_c$,  and
neutron radii, $r_n$,  for the nuclei used in Ref. \cite{Lalazissis97}
for determining the NL3 parameter set and in our fit, together with the
assumed error bars in percent.  The values obtained from the parameter
sets  SK272 and SK255 are shown in columns 5 and 6, respectively.  For the
sake of comparison, in Table \ref{f-nuclei} we also  give in columns 4 and
7 the results for  the NL3 and SGII interactions, respectively.  It is
evident from this table that the quality of our fit to the experimental
data is quite comparable to the results obtained with the NL3 and SGII
interactions.  We have also listed the values of $\Delta r = r_n -
r_p$, the difference between rms radii for neutrons and protons,
(not included in the fit). Experimental values for $\Delta r$  are
obtained  from the data for the neutron and charge radii and by using
$r_p=\sqrt{r_c^2-0.64}$. It is interesting to note that the values of
$\Delta r$ obtained for the SK272, SK255 and NL3 interactions are closer
and are quite large compared with the corresponding   results for the
SGII interaction.  This is consistent with the fact that the values of
the slope of the symmetry energy ($L$) associated with the SK272,
SK255 and NL3 interactions  are significantly larger than that associated with
the SGII interaction (see  Table \ref{n-matter}).  Since, as is well
known \cite{Friedman77}, the value of $\Delta r$ is sensitive to the
density dependent form adopted for the symmetry interaction.

\subsection{RPA results}
We have demonstrated very recently \cite{Agrawal03} that the strength
function for giant resonances are quite sensitive to the various numerical
approximations made. By numerical approximations we essentially mean, the
size of the box used for the discretization of the continuum, restriction
imposed on the maximum energy for the particle-hole excitations
($E_{ph}^{max}$) , smearing parameter ($\Gamma/2$) used to smear the
strength function, etc.  We have  shown in Ref. \cite{Agrawal03} that in
order to reproduce the results obtained in the continuum RPA calculation
the size of the box should be consistent with the value used for the
smearing width.  For $\Gamma/2=1$ MeV, one must use a large box of size
72 fm. In this work, we have used a box of size 90 fm and $\Gamma/2=1$
MeV. From Ref.  \cite{Agrawal03} we also note that in order to obtain
a  reliable value of the ISGMR centroid energy, $E_0$, accurate
within 0.1 MeV,   one must use $E_{ph}^{max} > 400$ MeV.  Here,
the centroid energy is given by  $E_0 = m_1/m_0$,  where $m_0$ and
$m_1$ are the non-energy-weighted and energy-weighted sums of $S(E)$
of Eq. (\ref{strength}), respectively.  In the present work, the
lowest value of $E_{ph}^{max}$ we have used is higher than 500 MeV. In
addition, the centroid energy depends strongly on the range adopted for
the excitation  energy interval of the giant resonance.  Nevertheless,
one often encounters in the published literature that the values of the
centroid energies are given without any reference to the corresponding
excitation energy range considered. For instance, in the case of the
$^{208}{\rm Pb}$  nucleus we find for the  SGII parameter set that $E_0
= $ 13.7, 13.9, 14.4 MeV  for the excitation energy  ranges $0 - 40$,
$0 - 60$ and $10 - 40$ MeV, respectively.  These differences are quite
significant,  since, as pointed  out earlier, a variation of $5\%$ in
$E_0$ corresponds to a change in $K_{nm}$ by $10\%$.  In what follows, we
shall concentrate mainly on the results for $E_0$ obtained by integrating
the strength function over the energy range $0 - 60$ MeV, since  the
RMF results presented in Ref. \cite{Giai01} for the NL3 parameter set
were obtained using  the same energy range  and the strength function
was smeared using $\Gamma/2 = 1$ MeV \cite{MaP}.

In Table \ref{isgmr-ecen} we give the results for  the ISGMR centroid
energy obtained using the parameter sets SK272 and SK255 and compare them
with the RRPA results of Ref. \cite{Giai01}  for the NL3  interaction. It
is clear from this table that the parameter set  SK272 yields values
 for  $E_0$ which are higher by about $4 - 6\%$  than those obtained using
the NL3 parameter set in the RRPA calculations. This implies that if we
reduce the compressibility  by about $10\%$, we can reproduce reasonably
well the RRPA results for the NL3 interaction. For this reason we have
generated another parameter set SK255 with $K_{nm}=255$ MeV, keeping
$J =  37.4$ MeV (see Table \ref{n-matter}).  In fact, we see that the
differences between the values of the  $E_0$ obtained from the parameter
sets SK255 and NL3 are on the level of the uncertainty associated with
the experimental data for $E_0$.  We emphasize that though the values
of  $K_{nm}$ associated with the SGII and SK255 parameter sets differ by
about 40 MeV, the values of $E_0$ for $^{208}{\rm Pb}$ nucleus for these
interactions are close within 0.3 MeV.  Thus, by  requiring  a fixed value of
$E_0$  we find  that an increase in $J$ by $10\%$ leads to an  increase in
$K_{nm}$ by about $5\%$.  We  also compute the values of $E_0$ over the
same energy range as used in experimental determination of the centroid
energy \cite{Youngblood99}.  It should be noted that the experimental
values of the energy range, given in the Table \ref{isgmr-ecen}, are
more or less symmetric around the corresponding $E_0$.  It  can be
seen from Table \ref{isgmr-ecen}, that for the parameter set SK255, we
obtain a good agreement with the experimental data for $E_0$, calculated
over the experimental excitation energy range.  We remark that, in our
calculations for the  $^{208}{\rm Pb}$ nucleus with the SK272 and SK255
parameter sets, the peak energy for the  isovector giant dipole resonance
is 13.2 and 13.3 MeV respectively, which is in good agreement with the
experimental value of $13.3\pm 0.1$ MeV \cite{Ritman93}.

\subsection{Conclusions}
In summary, we have analyzed  in detail the recent claim that the nuclear
matter incompressibility coefficient $K_{nm}$ extracted from the ISGMR
centroid energy  calculated using the relativistic and nonrelativistic
based RPA models differ by about $20\%$. For a meaningful comparison, we
have  generated  parameter sets for the  Skyrme interaction by a least
square  fitting procedure using exactly the same experimental data for
the bulk properties  of nuclei considered in Ref. \cite{Lalazissis97}
for determining  the NL3 parameterization of an effective Lagrangian used
in the relativistic mean field models.  Further, we also demanded in our
fitting procedure that the values of $K_{nm}$, $J$ and the charge radius
of the $^{208}{\rm Pb}$ nucleus should be very close to the results
obtained with  the NL3 interaction.  The parameter sets thus obtained
were used to calculate the ISGMR centroid energy for several nuclei.
For the parameter set SK272 ($K_{nm} = 272$ MeV), the calculated values
of  $E_0$    are higher by about $5\%$ compared to the corresponding
NL3 results. This implies that the difference in the value of $K_{nm}$
obtained in the relativistic and the nonrelativistic microscopic
models could be at most $10\%$.  In view of this, we generate another
parameter set  having $K_{nm} = 255$ MeV. As expected, the parameter
set associated with $K_{nm} = 255$ MeV, yields for  the ISGMR centroid
energies values which are quite close to the NL3 results.  Moreover, for
the SK255 parameter set, we find a good agreement with experimental data
for  $E_0$ for all the nuclei considered, provided, the corresponding
excitation energy ranges used in determining $E_0$ are the same as those
used in obtaining the experimental data.  We have thus shown that the
difference in the values of $K_{nm}$ obtained in the relativistic and
nonrelativistic models is mainly due to the differences  in the values
of the symmetry energy coefficient ($J$) and its slope ($L$) associated
with these models.

\newpage
\begin{table}
\caption{\label{parameters}Skyrme parameters for different interactions
used in the present calculations.  Value of the  parameters for the SGII
interaction are taken from Ref. \cite{Giai81}.}
\begin{tabular}{|cccc|}
\hline
Parameter&  SK272& SK255&SGII\\
\hline
 $t_0$(MeV $\cdot$ fm$^{3}$)& -1496.84&-1689.35&-2645\\
 $t_1$(MeV $\cdot$ fm$^{5}$)&   397.66&  389.30&  340\\
 $t_2$(MeV $\cdot$ fm$^5$)&  -112.82& -126.07&  -41.9\\
 $t_3$(MeV $\cdot$ fm$^{3(1+\alpha)}$) &10191.64&10989.60&15595\\
 $x_0$&    0.0008&   -0.1461&    0.09\\
 $x_1$&    0.0102&    0.1160&   -0.0588\\
 $x_2$&    0.0020&    0.0012&    1.425\\
 $x_3$&   -0.5519&   -0.7449&    0.06044\\
 $alpha$&    0.4492&    0.3563&   1/6\\
 $W_0$(MeV $\cdot$ fm$^5$)&   106.58&   95.39&  105\\
\hline
\end{tabular}
\end{table}
\newpage
\begin{table}
\caption{\label{n-matter}Nuclear matter properties calculated from RMF
theory with the NL3 parameter set and the nonrelativistic HF calculations
with different Skyrme parameter sets. The "experimental data" are the ones
used in Ref. \cite{Lalazissis97}  in the least square fit  together with
the bulk properties for finite nuclei in obtaining the NL3 parameter set.
The values in parenthesis represent the error bars (in percent)  used
in the fit.}
\begin{tabular}{|cccccc|}
\hline
 & Exp. & NL3& SK272& SK255& SGII\\
\hline
$E/A$(MeV) &-16.0(5)   & -16.299& -16.280& -16.334& -15.67\\
$K_{nm}$ (MeV)  &250.0(10)& 271.76& 271.55& 254.96& 214.57\\
$\rho_0$ (fm$^{-3}$)&0.153(10)  & 0.148& 0.155& 0.157& 0.159\\
$m^*/m$&  & 0.60& 0.77& 0.80& 0.79\\
$J$ (MeV)&33.0(10)  & 37.4& 37.4&  37.4& 26.8\\
$L$ (MeV)&  & 118.5& 91.7& 95.0& 37.6\\
\hline
\end{tabular}
\end{table}
\newpage
\begin{table}
\caption{\label{f-nuclei} Experimental data for the total binding energy
($E$ ) in MeV, charge  ($r_c$) and neutron ($r_n$) radii in fm for the
nuclei used in the least square fitting procedure.  The parameter sets
thus obtained are named as SK272 and SK255. For comparison we also give
the results obtained from the NL3 and SGII interactions.  The values
in parenthesis represent the error bars (in percent)  used in the fit.
The quantity $\Delta r = r_n - r_p$ (not included in the fit) is the
difference between the radii for the neutrons and protons. }

\begin{tabular}{|ccccccc|}
\hline
Nucleus&Property& Exp. & NL3& SK272& SK255& SGII\\
\hline
$^{16}{\rm O}$&$E$& -127.62(0.1)& -128.83& -127.76& -128.05& -131.93\\
&$r_c$& 2.730(0.2)& 2.730& 2.800& 2.813& 2.793\\
&$r_n$&  & 2.580& 2.662& 2.674& 2.650\\
\hline
$^{40}{\rm Ca}$&$E$& -342.06(0.1)& -342.02& -341.35& -342.50& -342.42\\
&$r_c$&3.450(0.2)& 3.469& 3.496& 3.504& 3.490\\
&$r_n$& 3.370(2.0)& 3.328& 3.363& 3.369& 3.348\\
&$\Delta r$&0.014   & -0.047& -0.041& -0.043& -0.049\\
\hline
$^{48}{\rm Ca}$&$E$&  -416.00(0.1)& -415.15& -414.17& -413.89& -418.22\\
&$r_c$& 3.451(0.2)& 3.470& 3.524& 3.531& 3.526\\
&$r_n$& 3.625(2.0)& 3.603& 3.635& 3.649& 3.582\\
&$\Delta r$&0.268 & 0.227& 0.203& 0.210& 0.147\\
\hline
$^{90}{\rm Zr}$&$E$&  -783.90(0.1)& -782.63& -782.73& -783.28& -775.49\\
&$r_c$&  4.258(0.2)& 4.287& 4.282& 4.286& 4.286\\
&$r_n$& 4.289(2.0)& 4.306& 4.310& 4.317& 4.266\\
&$\Delta r$&0.107  & 0.094& 0.103& 0.106& 0.056\\
\hline
$^{116}{\rm Sn}$&$E$&   -988.69(0.1)& -987.67& -982.37& -984.48& -971.66\\
&$r_c$&  4.627(0.2)& 4.611& 4.617& 4.619& 4.630\\
&$r_n$& 4.692(2.0)& 4.735& 4.696& 4.701& 4.639\\
&$\Delta r$&0.135   & 0.194& 0.149& 0.152& 0.079\\
\hline
$^{132}{\rm Sn}$&$E$&    -1102.90(0.1)& -1105.44& -1097.36& 1100.04&-1105.17\\
&$r_c$&   & 4.709& 4.725& 4.726& 4.735\\
&$r_n$&  & 4.985& 4.964& 4.975& 4.867\\
\hline
$^{208}{\rm Pb}$&$E$&    -1636.47(0.1)& -1639.54& -1631.78& -1637.48& -1622.21\\
&$r_c$&    5.503(0.2)& 5.520& 5.503& 5.503& 5.519\\
&$r_n$&   5.593(2.0)& 5.741& 5.687& 5.694& 5.597\\
&$\Delta r$& 0.148   & 0.279& 0.243& 0.250& 0.136\\
\hline
\end{tabular}
\end{table}
\newpage
\begin{table}
\caption{\label{isgmr-ecen}ISGMR centroid energy $E_0=m_1/m_0$  (in MeV) obtained by
integrating over the energy range  $\omega_{1} - \omega_{2}$ MeV with the strength
function smeared by using $\Gamma/2 = 1$ MeV. The experimental data is taken from Ref.
\cite{Youngblood99}.}
\begin{tabular}{|ccccccc|}
\hline
Nucleus &  $\omega_{1} - \omega_{2}$&Exp. & NL3& SK272& SK255& SGII\\
\hline
$^{90}{\rm Zr}$&$0 - 60$&    &18.7& 20.0  &18.9 & 18.3\\
&$10 - 26$& 17.89$\pm$0.20&   & 19.3 & 18.4 & 17.9\\
\hline
$^{116}{\rm Sn}$&$0 - 60$&  &17.1& 18.0 & 17.5 & 16.6\\
&$10 - 23$&16.07$\pm$0.12&  &17.4 & 16.9& 16.3\\ 
\hline
$^{144}{\rm Sm}$&$0 - 60$&  &16.1& 17.1 & 16.4 & 15.6\\
&$10 - 22$&15.39$\pm$0.28 &&16.5& 15.9& 15.2\\
\hline
$^{208}{\rm Pb}$&$0 - 60$&   &14.2& 14.7 & 14.2 &13.9\\ 
&$8 - 21$&14.17$\pm$0.28&   &14.2 &13.8&13.6\\
\hline
\end{tabular}
\end{table}
\end{document}